\documentclass[11pt]{article}

\usepackage[margin=1in]{geometry}
\usepackage{iftex}
\usepackage{amsmath,mathtools}
\ifPDFTeX
  \usepackage[T1]{fontenc}
  \usepackage[utf8]{inputenc}
  \usepackage{textcomp}
  \usepackage{amssymb}
\else
  \usepackage{unicode-math}
  \defaultfontfeatures{Scale=MatchLowercase}
  \defaultfontfeatures[\rmfamily]{Ligatures=TeX,Scale=1}
\fi
\usepackage{lmodern}
\usepackage{xcolor}
\IfFileExists{microtype.sty}{%
  \usepackage[]{microtype}
  \UseMicrotypeSet[protrusion]{basicmath}
}{}
\usepackage{booktabs,longtable,array}

\usepackage{calc}
\usepackage{etoolbox}
\makeatletter
\patchcmd\longtable{\par}{\if@noskipsec\mbox{}\fi\par}{}{}
\makeatother
\usepackage{graphicx}
\usepackage{float}
\IfFileExists{xurl.sty}{\usepackage{xurl}}{}
\usepackage{hyperref}
\hypersetup{
  colorlinks=true,
  linkcolor=blue,
  urlcolor=blue,
  citecolor=blue,
  pdftitle={The Engineering of Skew},
  pdfauthor={Gregory A. Fanous},
  pdfsubject={Path-dependent asymmetric volatility management},
  pdfkeywords={skewness, drawdown, recovery, volatility
management, portfolio construction, path-dependent risk}
}

\title{The Engineering of Skew\\[0.35em]{\large A Path-Dependent
Framework for Asymmetric Volatility Management}}
\author{Gregory A. Fanous}
\date{May 2026}

\begin{document}

\maketitle

\begin{abstract}
Volatility is the language in which finance often describes risk, but it
is not the language in which institutions experience risk. Allocators
live through drawdowns, liquidity needs, spending rules, rebalance
decisions, board oversight, and the interval between a prior high-water
mark and full recovery. This paper develops a path-dependent framework
for asymmetric volatility management. The arithmetic of recovery is
nonlinear: after a drawdown of depth \(D\), the required gain is
\(R=\frac{1}{1-D}-1\). Lower volatility can improve geometric
compounding through the familiar small-return approximation
\(g \approx \mu-\frac{1}{2}\sigma^2\), but symmetric de-risking can also
impair recovery if it sacrifices too much upside participation. The
relevant design problem is therefore not volatility reduction in
isolation; it is conditional exposure shaping. Skew engineering is
defined here as the portfolio construction discipline of reducing
harmful downside participation more than productive upside
participation, controlling submergence, and preserving enough recovery
participation to sustain compounding under adverse regimes. The
resulting Recovery-Efficiency Protocol links drawdown depth, time
underwater, recovery burden reduction, and rebound participation into an
allocator-facing reporting discipline. Machine learning and AI methods
are framed as tools for conditional estimation, regime mapping,
robustness testing, and model-risk governance, not as market prediction.
\end{abstract}

\section{Contributions and Scope}\label{contributions-and-scope}

This paper develops a practical framework for portfolio design. Its
central premise is simple: institutions do not experience risk as an
abstract distribution. They experience it as a path through time, with
spending needs, governance pressure, liquidity constraints, and the
recurring question of whether capital can recover quickly enough to
remain useful.

The paper makes four contributions. First, it frames the return path,
rather than volatility alone, as the object of portfolio design. Second,
it defines recovery burden reduction as a practical diagnostic for
comparing how different paths change the gain required to recover from
loss. Third, it places upside and downside capture inside a
path-dependent framework for interpreting skew. Fourth, it names a
Recovery-Efficiency Protocol that joins drawdown depth, time underwater,
recovery burden, and rebound participation into a falsifiable reporting
discipline.

The paper is intentionally not an empirical backtest, product
description, or claim of persistent alpha. It does not propose a
universal optimal allocation rule. Its purpose is narrower and
practical: to clarify the measurement and governance problem faced by
institutions that seek lower left-tail participation without fully
abandoning upside participation.

\section{The Institutional Problem Is Not Volatility
Alone}\label{the-institutional-problem-is-not-volatility-alone}

Volatility is a useful abstraction. It compresses a return path into a
single measure of dispersion, usually expressed as the standard
deviation of periodic returns, \(\sigma\). That compression has value:
it permits comparison across portfolios, enables optimization, and
provides a common vocabulary for risk budgeting. Mean-variance portfolio
theory begins from precisely this tradeoff between expected return and
variance {[}Markowitz, 1952{]}. But volatility is incomplete as an
institutional risk measure because it treats upside and downside
deviations symmetrically.

The lived experience of institutional risk is not symmetric in that way.
Fiduciaries rarely convene emergency meetings because a portfolio
compounded too quickly. They respond to drawdown, liquidity pressure,
tracking stress, contribution uncertainty, and the possibility that a
portfolio will remain below its prior peak long enough to impair the
institution's mission. A conventional variance measure can describe the
roughness of the ride, but it does not fully describe whether the ride
forces a sale, triggers a mandate review, violates a risk limit, or
prevents a portfolio from funding obligations. This is why drawdown and
recovery often dominate the governance experience of risk {[}Van Hemert
et al., 2020; Rook et al., 2023{]}.

The distinction between path-independent and path-dependent risk is
central. A path-independent metric evaluates a distribution without
regard to the sequence of returns. A path-dependent metric asks how
capital moves through time. The sequence matters because compounding is
multiplicative:

\[
W_T = W_0 \prod_{t=1}^{T}(1+r_t).
\]

With fixed exposures and no external cash flows, the order of a given
set of simple returns does not change terminal wealth. Institutions
rarely live in that clean laboratory. Spending, contributions,
rebalancing rules, mandate limits, liquidity needs, and risk reductions
after loss can make sequence economically decisive. Even before those
frictions are introduced, sequence changes the lived path: drawdown
depth, time underwater, governance stress, and the ability to rebalance
from strength rather than weakness. Volatility is therefore one input to
a larger path problem, not a complete description of institutional risk.

Some volatility is productive: compensated fluctuation that preserves or
enhances long-term compounding. It may include upside convexity,
tactical participation in favorable regimes, or temporary mark-to-market
movement that does not deepen or prolong submergence. Other volatility
is harmful because it appears when capital is scarce, liquidity is
valuable, correlations are rising, and investor behavior is vulnerable.
This kind of volatility deepens drawdowns, lengthens recovery, and
reduces the portfolio's ability to rebalance into opportunity.

\section{The Arithmetic of Recovery}\label{the-arithmetic-of-recovery}

If a portfolio declines by a fraction \(D \in [0,1)\) from its prior
peak, the gain required to return to that peak is

\[
R(D) = \frac{1}{1-D}-1.
\]

This equation is arithmetic rather than a model assumption. A \(10\%\)
drawdown requires an \(11.1\%\) recovery return. A \(25\%\) drawdown
requires \(33.3\%\). A \(50\%\) drawdown requires \(100\%\). As \(D\)
approaches \(1\), the required recovery return grows without bound. The
nonlinearity is the central mathematical reason that avoiding deep
losses can be more valuable than merely seeking high average gains
{[}Mauboussin and Callahan, 2025{]}.

\begin{table}[H]
\centering
\small
\caption{Drawdown depth and required recovery return}
\begin{tabular}{rrr}
\toprule
Drawdown depth $D$ & Portfolio value after drawdown & Required recovery return $R(D)$ \\
\midrule
$5\%$ & $0.95$ & $5.3\%$ \\
$10\%$ & $0.90$ & $11.1\%$ \\
$15\%$ & $0.85$ & $17.6\%$ \\
$20\%$ & $0.80$ & $25.0\%$ \\
$25\%$ & $0.75$ & $33.3\%$ \\
$30\%$ & $0.70$ & $42.9\%$ \\
$40\%$ & $0.60$ & $66.7\%$ \\
$50\%$ & $0.50$ & $100.0\%$ \\
$60\%$ & $0.40$ & $150.0\%$ \\
$80\%$ & $0.20$ & $400.0\%$ \\
\bottomrule
\end{tabular}
\end{table}

Drawdowns are unavoidable in risk assets, so the relevant point is not
avoidance. It is that drawdown depth is not linearly related to recovery
burden. A portfolio that loses \(20\%\) must gain \(25\%\) to recover. A
portfolio that loses \(40\%\) must gain \(66.7\%\). The second loss is
twice as deep but requires more than two and a half times the recovery
return.

For a benchmark drawdown episode \(k\), let \(a_k\) denote the benchmark
high-water-mark date at the beginning of the episode, \(m_k\) the
benchmark trough date, and \(b_k\) the date when the benchmark returns
to its prior high. Using the benchmark episode as the common calendar
window, define

\[
D_k^B = 1-\min_{a_k\leq t\leq b_k}\frac{W_t^B}{W_{a_k}^B},
\]

and

\[
D_{k|B}^P = \max\left(0,\ 1-\min_{a_k\leq t\leq b_k}\frac{W_t^P}{W_{a_k}^P}\right).
\]

The notation \(D_{k|B}^P\) emphasizes that the portfolio drawdown is
measured over the benchmark-defined episode. A simple diagnostic for
recovery burden reduction is

\[
BR_k = 1-\frac{R(D_{k|B}^P)}{R(D_k^B)},
\]

defined when \(D_k^B>0\). Positive \(BR_k\) means the portfolio reduced
the required recovery gain relative to the benchmark event; negative
\(BR_k\) means it increased the burden. This diagnostic should be
interpreted with recovery participation and time underwater, not alone.
A portfolio that reduces drawdown but eliminates upside participation
may still recover slowly.

\section{Volatility Drag and the Geometry of
Compounding}\label{volatility-drag-and-the-geometry-of-compounding}

Investors experience geometric returns, not arithmetic averages. The
arithmetic mean return is

\[
\mu = \frac{1}{T}\sum_{t=1}^{T} r_t,
\]

while the realized geometric return is

\[
g = \left(\prod_{t=1}^{T}(1+r_t)\right)^{1/T}-1.
\]

For small simple returns, a second-order approximation to expected log
growth gives the familiar relationship

\[
\mathbb{E}[\log(1+r)] \approx \mu-\frac{1}{2}\sigma^2.
\]

This is the sense in which volatility can create a variance penalty for
compound growth. The approximation is not a universal identity, and it
should not be applied mechanically to large returns, leverage,
fat-tailed distributions, or time-varying exposures. Its value is
conceptual: for a given arithmetic return, variance reduces expected log
growth. The same intuition underlies the Kelly criterion literature,
though institutional portfolios generally cannot treat long-run growth
maximization as the only objective because drawdown, liquidity, and
governance constraints bind {[}Kelly, 1956{]}.

The approximation also explains why volatility reduction alone is
insufficient. A strategy that reduces \(\sigma\) but sacrifices too much
\(\mu\) can reduce geometric return. If the variance penalty falls by
\(1\%\) but the arithmetic return falls by \(3\%\), the investor may be
worse off. The design problem is therefore to reduce volatility where it
is most damaging while preserving enough participation in
return-producing states.

This is where the distinction between harmful and productive volatility
becomes operational. A portfolio that suppresses upside variation while
leaving exposure to left-tail states, liquidity events, and correlation
breakdowns intact has not solved the institutional problem. It has
merely made the return stream look smoother until smoothness is most
needed.

The low-volatility evidence shows both promise and limitation. The
low-volatility anomaly suggests that lower-risk equities have
historically delivered attractive risk-adjusted outcomes relative to
higher-volatility equities, challenging the simplified view that higher
risk must always produce higher return {[}Baker et al., 2011{]}. But
implementation matters. A naive low-volatility portfolio can become a
concentrated bet on defensive sectors, interest-rate sensitivity, or
crowded factors. A more robust approach asks whether lower volatility
comes from genuine improvement in compounding efficiency or from hidden
return sacrifice and unrecognized exposures {[}Dekhayser et al., 2025;
Xiang et al., 2023{]}.

\section{The Failure of Symmetric
De-Risking}\label{the-failure-of-symmetric-de-risking}

If the only objective were lower volatility, the solution would be
simple: hold more cash, reduce beta, or scale every exposure down
proportionally. Such symmetric de-risking reduces upside and downside
together. It lowers the amplitude of the return stream, but it does not
necessarily improve recovery dynamics. In some cases, it can make
recovery slower because the portfolio has less ability to participate
after a decline.

This is the failure of symmetric de-risking: it treats exposure as the
problem rather than conditional exposure as the design variable. The
issue is not simply how much risk the portfolio takes on average. The
issue is when it takes risk, what type of risk it takes, and how that
risk behaves across regimes. A portfolio with \(70\%\) upside capture
and \(70\%\) downside capture is lower beta, but not necessarily more
intelligently shaped. A portfolio with \(85\%\) upside capture and
\(60\%\) downside capture has a different profile. It participates
meaningfully when markets rise while reducing more of the damage when
markets fall.

Traditional risk budgeting allocates capital or volatility across asset
classes, strategies, or factors. Asymmetric volatility budgeting asks a
conditional question: how much of the volatility budget is spent in
states that help compounding versus states that impair it? In practice,
this may involve reducing exposure when downside volatility and
correlation risk are rising, adding diversifiers whose payoff is
expected to improve in stress, or using convex instruments that reshape
the left tail. It also requires recognizing tradeoffs: protection has
costs, signals can fail, and models are subject to estimation error.

Volatility-managed portfolios provide one empirical and theoretical
bridge from static allocation to conditional exposure management.
Scaling exposure by recent volatility can improve risk-adjusted
performance in some settings, but that evidence should be treated as
support for conditional risk timing rather than a complete
implementation rule {[}Moreira and Muir, 2017{]}. The broader takeaway
is that the timing of risk exposure is itself a portfolio design
variable.

\section{Asymmetric Capture as a Practical Measure of
Skew}\label{asymmetric-capture-as-a-practical-measure-of-skew}

Statistical skewness is important, but many allocators need a measure
that is intuitive, communicable, and tied to portfolio experience.
Upside and downside capture ratios provide that bridge. They translate
the abstract idea of skew into a practical question: how much of the
benchmark's wealth creation does the portfolio retain, and how much of
the benchmark's wealth destruction does it absorb?

Let \(B_t\) denote benchmark return, \(P_t\) denote portfolio return,
\(\mathcal{S}^{+}=\{t:B_t>0\}\) denote benchmark-positive periods,
\(\mathcal{S}^{-}=\{t:B_t<0\}\) denote benchmark-negative periods, and
\(n_+\) and \(n_-\) denote the number of observations in those sets. For
headline reporting, this paper uses conditional geometric-period
capture:

\[
UC_g =
\frac{\left(\prod_{t\in\mathcal{S}^{+}}(1+P_t)\right)^{1/n_+}-1}
{\left(\prod_{t\in\mathcal{S}^{+}}(1+B_t)\right)^{1/n_+}-1},
\]

and

\[
DC_g =
\frac{\left(\prod_{t\in\mathcal{S}^{-}}(1+P_t)\right)^{1/n_-}-1}
{\left(\prod_{t\in\mathcal{S}^{-}}(1+B_t)\right)^{1/n_-}-1},
\]

where the relevant benchmark conditional return is nonzero.\footnote{Arithmetic
  capture, based on conditional average returns, is also common and
  often intuitive. The product-minus-one convention measures percentage
  capture of conditional cumulative gains or losses; it can be useful as
  a conditional wealth companion, but over long samples it may be
  dominated by compounding and observation counts. A pure growth-factor
  ratio,
  \(\widetilde{C}(\mathcal{S})=\frac{\prod_{t\in\mathcal{S}}(1+P_t)}{\prod_{t\in\mathcal{S}}(1+B_t)}\),
  should be named separately. Ratios become unstable when the benchmark
  conditional return is close to zero, so the underlying conditional
  returns and observation counts should be reported in those cases.}
This convention asks how much of the benchmark's typical compounded
positive-period and negative-period experience the portfolio captured.
Higher upside capture is generally desirable, all else equal. Lower
positive downside capture is generally desirable; if \(DC_g<0\), the
portfolio gained during benchmark-negative periods.

The practical skew diagnostic is not either capture ratio alone, but the
relationship between them:

\[
\mathcal{A}_{cap}=UC_g-DC_g.
\]

Positive capture asymmetry indicates that the portfolio participated
more in benchmark-positive states than in benchmark-negative states. A
portfolio with \(70\%\) upside capture and \(70\%\) downside capture is
lower beta, but has no capture asymmetry. A portfolio with \(85\%\)
upside capture and \(60\%\) downside capture preserves materially more
upside participation than downside participation.

This paper's contribution is not the invention of capture ratios
themselves; those are established reporting measures {[}Pak, 2014{]}.
The contribution is to place them inside a path-dependent framework for
skew. Statistical skewness describes the shape of a return distribution;
asymmetric capture describes the conditional shape of the realized
wealth path. It asks whether the portfolio reduces harmful market
participation more than productive market participation.

The connection to recovery speed is direct. Lower downside capture can
reduce drawdown depth, while adequate upside capture helps the portfolio
participate in the rebound. Because required recovery \(R(D)\) is convex
in drawdown depth, reducing downside capture in severe regimes can have
a nonlinear effect on the recovery burden. Thus asymmetric capture turns
the arithmetic of recovery into an observable performance diagnostic.

Asymmetric capture is not a complete measure of skew. It depends on
benchmark choice, horizon, fees, market regime, and the distribution of
up and down periods. Its value is that it forces the right conversation:
the objective is not merely to reduce \(\sigma\), but to improve the
conditional relationship between the portfolio and the market states
that matter most.

\section{Drawdown, Recovery, and
Submergence}\label{drawdown-recovery-and-submergence}

Let the high-water mark process be

\[
H_t = \max_{0\leq s\leq t} W_s,
\]

and define drawdown at time \(t\) as

\[
DD_t = 1-\frac{W_t}{H_t}.
\]

A submergence episode begins when \(W_t < H_t\) and ends when
\(W_t \geq H_t\) again. The maximum drawdown within the episode captures
depth. The duration of the episode captures time underwater. Together,
drawdown and recovery describe the path-dependent investor experience
more directly than volatility alone {[}Rook et al., 2023{]}.

Submergence is closer to institutional pain than volatility because it
captures both magnitude and duration. A short, shallow drawdown may be
uncomfortable but manageable. A deep drawdown that recovers quickly may
still be tolerable for some institutions if liquidity is available and
governance remains stable. A prolonged submergence, however, can create
compounding damage even if annualized volatility later appears moderate.
It can force distributions from a depressed capital base, reduce funded
status, impair strategic rebalancing, and create pressure to terminate
managers near troughs.

For drawdown episode \(k\), a compact recovery-efficiency profile can be
written as

\[
\mathcal{P}_{k|B} = \left(D_{k|B}^P,\ U_{k|B}^P,\ BR_k,\ UC_k^{rec}\right),
\]

where \(D_{k|B}^P\) is portfolio drawdown depth over the
benchmark-defined episode, \(U_{k|B}^P\) is the portfolio's time below
its value at the benchmark episode start, \(BR_k\) is recovery burden
reduction relative to the benchmark, and \(UC_k^{rec}\) is upside
capture during the benchmark's recovery interval. This tuple is not a
universal utility function. It is a reporting discipline: recovery
evaluation should jointly examine loss depth, time below the relevant
high-water mark, burden reduction, and participation in the subsequent
rebound.

This framing also clarifies why aggressive upside exposure alone is not
enough. A portfolio that captures recoveries but suffers large drawdowns
may require too much rebound to recover efficiently. Conversely, a
portfolio that avoids losses but captures little upside may remain
underwater for too long. The design target is the joint outcome, not any
single statistic.

\section{The Recovery-Efficiency
Protocol}\label{the-recovery-efficiency-protocol}

The framework becomes useful only when it changes the questions asked in
an investment memo. The Recovery-Efficiency Protocol is the practical
interface for doing that. It does not collapse a strategy into one
score. It creates a consistent table of evidence so that an allocator
can see whether a design reduced the recovery burden, preserved rebound
participation, and avoided creating its own long submergence.

First, state the protocol inputs: benchmark, candidate strategy, return
frequency, sample period, fee and data treatment, liquidity assumptions,
and the benchmark drawdown threshold used to define episodes. Capture
ratios, drawdowns, and recovery statistics are all shaped by these
choices. They should be visible rather than hidden in the machinery.

Second, report the full-period profile: annualized return, annualized
volatility, maximum drawdown, longest time underwater, upside capture,
downside capture, and capture asymmetry. This profile separates ordinary
performance from path shape. A strategy that improves capture asymmetry
but suffers a long independent submergence has not solved the
institutional problem.

Third, segment benchmark drawdown episodes. For each episode, report the
benchmark peak, trough, and recovery dates; benchmark drawdown depth;
portfolio drawdown over the same benchmark-defined window; recovery
burden reduction \(BR_k\); portfolio underwater duration; and portfolio
upside capture during the benchmark's recovery interval, \(UC_k^{rec}\).
This makes the central question observable: did the strategy preserve
capital when recovery became expensive, and did it still participate
when recovery began?

Fourth, report the portfolio's own submergence episodes separately. A
strategy can look excellent inside benchmark windows while creating its
own long underwater periods. The benchmark-event view and portfolio-own
view answer different questions, and both are needed.

Fifth, attach the cost of the design. Protection may require option
premium, turnover, tracking error, lower average beta, leverage
constraints, or a willingness to look wrong for long periods. A strategy
earns confidence not by hiding these costs, but by making them explicit
enough for an institution to govern.

Finally, use the protocol before proposing construction changes. Define
the recovery constraint, compare symmetric de-risking with asymmetric
candidates, and report the tradeoff budget. The protocol is not an alpha
model or a universal allocation rule. It is a discipline for deciding
whether a portfolio's path behavior improved in the dimensions that
institutions actually experience.

\section{Skew, Coskewness, and Portfolio-Level
Utility}\label{skew-coskewness-and-portfolio-level-utility}

Skewness is the standardized third central moment of a return
distribution. Positive skewness indicates a distribution with a longer
or fatter right tail; negative skewness indicates a longer or fatter
left tail. Investors generally prefer positive skewness because it
implies greater potential for large favorable outcomes, and they
generally dislike negative skewness because it implies exposure to large
adverse outcomes. But standalone skewness is not enough. What matters
for a diversified portfolio is how an asset behaves when the rest of the
portfolio is under stress.

This is the concept of coskewness. In the same way that covariance
measures how an asset moves with the market in average conditions,
coskewness measures how it contributes to portfolio asymmetry,
especially in extreme states. Harvey and Siddique's work on conditional
skewness showed that systematic skewness can be a priced component of
expected return {[}Harvey and Siddique, 1999; Harvey and Siddique,
2000{]}. Assets that worsen portfolio outcomes during market stress may
require compensation; assets that improve stress behavior may be
accepted at lower expected return because they provide insurance-like
value.

Conditional skewness is also important. The skewness of returns is not
constant through time. Markets move through regimes of calm,
acceleration, crowding, liquidity stress, and deleveraging. A strategy
may appear diversifying in ordinary periods and still exhibit negative
coskewness in crisis. For an allocator, this is a central failure mode.
Diversification that disappears precisely when needed is not useless,
but it is less valuable than average correlations imply.

Portfolio-level utility is therefore shaped by state dependence. If an
asset's payoff is positive or resilient when \(W_t\) is far below
\(H_t\), it may help reduce submergence and improve the
recovery-efficiency profile. If an asset's payoff is most negative when
\(DD_t\) is already high, it can deepen harmful volatility. This is the
bridge between higher-moment finance and practical institutional design.
The relevant skew is not merely statistical. It is conditional,
contextual, and tied to the recovery path.

This also connects asymmetric volatility management to tail-risk
optimization. Conditional value-at-risk and related expected-shortfall
methods provide useful ways to optimize or hedge portfolios against
adverse tail outcomes {[}Rockafellar and Uryasev, 2000{]}. Those methods
are not substitutes for path-aware reporting, but they are useful
complements because the left tail and the recovery path are economically
linked.

\section{Engineering Skew as a Portfolio Design
Problem}\label{engineering-skew-as-a-portfolio-design-problem}

This paper defines skew engineering as the deliberate portfolio
construction process of shaping conditional return behavior so that the
portfolio seeks to reduce harmful downside participation more than
productive upside participation, improve recovery-efficiency
diagnostics, and control submergence, subject to cost, liquidity,
governance, and model-risk constraints.

Several methods can contribute to skew engineering. Dynamic allocation
can reduce exposure when risk conditions deteriorate and restore
exposure when conditions improve. This may be based on volatility,
trend, valuation, liquidity, macro conditions, or ensemble signals. The
benefit is adaptability. The tradeoff is model risk, turnover, and the
possibility of false signals.

Volatility targeting can scale exposure to maintain a more stable level
of portfolio risk. When volatility rises, exposure is reduced; when
volatility falls, exposure can be increased. Moreira and Muir show that
volatility-managed portfolios can improve risk-adjusted performance in
some settings {[}Moreira and Muir, 2017{]}. In this framework, the
implementation question is whether the re-risking rule preserves
recovery participation rather than merely stabilizing ex ante
volatility.

Trend and risk-off signals can help portfolios respond to sustained
changes in market direction. Managed futures and trend-following
strategies have documented evidence across long histories and have often
performed well during prolonged crisis environments {[}Hurst et al.,
2017; Kaminski and Zhao, 2023{]}. Their advantage is that they can move
long or short across asset classes and may generate crisis-alpha-like
behavior during sustained trends. Their cost is whipsaw: in choppy or
rapidly reversing markets, trend systems can lose money on both sides of
the move.

Cross-asset diversification remains important, but it must be evaluated
conditionally. Government bonds, currencies, commodities,
inflation-sensitive assets, and alternative risk premia may contribute
differently across regimes. The design question is not simply average
correlation; it is stress correlation and coskewness. If an asset
diversifies in normal times but sells off with equities during liquidity
stress, its diversification value is overstated by unconditional
measures.

Option overlays can explicitly reshape the payoff distribution.
Protective puts and long-volatility components can truncate downside or
provide direct exposure to volatility spikes. These designs may improve
asymmetry, but premium cost, implied-volatility path dependence,
liquidity, counterparty exposure, and implementation discipline are
material {[}Driscoll and Giordano, 2024; Gambhirwala and Brundage,
2021{]}.

Skew engineering is therefore not a single product or technique. It is a
design philosophy and reporting discipline. It asks how each component
affects the left tail, the right tail, recovery participation, liquidity
during stress, and the institution's ability to stay invested. It
recognizes that protection has a cost, every hedge can fail, and
implementation quality matters as much as concept.

\section{AI, Conditional Estimation, and Model-Risk
Governance}\label{ai-conditional-estimation-and-model-risk-governance}

Machine learning and AI methods should be framed carefully in portfolio
construction. Their credible role is conditional estimation under
uncertainty, not point prediction with certainty. Markets are noisy,
adaptive, and reflexive. No model can eliminate uncertainty, and no
model should be presented as able to predict adverse regimes reliably.
This is not just philosophical caution. Financial machine-learning work
is vulnerable to overfitting, repeated testing, unstable regimes, and
fragile labels; current interagency model-risk guidance also emphasizes
model development and use, validation and monitoring, governance,
controls, and a risk-based approach to model oversight {[}Bailey et al.,
2017; López de Prado, 2018; Federal Reserve Board, FDIC, and OCC,
2026{]}.

Traditional linear models often assume stable relationships. They can be
powerful, but they may struggle when relationships are nonlinear or
conditional. For example, the relationship between volatility and future
drawdown may differ across liquidity regimes. Trend signals may behave
differently in a slow bear market than in a rapid correction.
Correlations across asset classes may behave differently in inflation
shocks than in deflation shocks. Machine-learning methods can help map
these conditional patterns, especially when used in ensembles and
constrained by economic logic.

The potential value is not a single forecast. It is conditional mapping:
estimating when the same exposure may have different implications for
drawdown, recovery, or coskewness. It is robustness: testing whether
signals survive changes in sample period, market definition, transaction
cost assumption, and stress scenario. It is adaptation: updating
estimates as market structure changes while avoiding excessive turnover.

The governance language matters. An AI-enabled portfolio process should
be able to explain what data it uses, what signals it emphasizes, how it
avoids overfitting, how it behaves under stress, and what risks remain.
It should emphasize stress behavior rather than point forecasts. It
should define the conditions under which exposure is reduced, restored,
or diversified. It should be evaluated not only by backtested return,
but by drawdown, time underwater, downside capture, upside capture,
turnover, liquidity, and failure modes.

In this framework, AI supports skew engineering only when it is
subordinate to risk controls. It may improve conditional estimates, but
liquidity discipline, diversification, stress testing, and fiduciary
clarity remain primary. The proper question is not whether a model can
predict the market. The proper question is whether a model improves the
portfolio's conditional exposure map without introducing hidden
fragility.

\section{Institutional Recovery
Speed}\label{institutional-recovery-speed}

Institutional recovery speed is a design objective, not a guarantee. It
translates the preceding diagnostics into a governance question: is the
portfolio built to preserve enough capital through adverse regimes,
participate enough in subsequent recoveries, and reduce the depth and
duration of submergence? It is not a promise that recovery will occur by
a particular date or that losses will be avoided. It is a way to
evaluate whether the return path is aligned with institutional
constraints.

The objective is best evaluated as a profile rather than a point
estimate. Maximum drawdown, time underwater, recovery burden reduction,
and recovery participation each describe a different part of the path. A
strategy can look successful on one dimension while failing another: it
may soften losses but lag the rebound, or it may capture the rebound
only after imposing a recovery burden too large for the institution to
tolerate.

Institutional recovery speed also requires humility about costs. Option
overlays may bleed premium. Trend systems may whipsaw. Volatility
targeting may lag rebounds. Diversifiers may fail in liquidity crises.
AI models may overfit. The point is not to deny these tradeoffs. It is
to budget for them explicitly and decide which costs are acceptable in
exchange for improved path behavior.

This perspective changes how portfolios are evaluated. Instead of asking
only, ``What is the expected return and volatility?'' the allocator
asks: What is the drawdown profile? What is the likely time underwater?
What is the downside capture? How much upside capture remains after
protection costs? How does the strategy behave when correlations rise?
What are the liquidity requirements? What assumptions fail under stress?
How does the process adapt without becoming unstable?

When those questions are answered clearly, volatility reduction becomes
more than a cosmetic reduction in standard deviation. It becomes part of
a path-aware design discipline in which the task is engineered
asymmetry.

\section{Interpretation and
Boundaries}\label{interpretation-and-boundaries}

This paper is conceptual, but the framework is meant to be tested rather
than protected from testing. It does not estimate a persistent skew
premium, prove that any particular implementation will outperform, or
claim that recovery can be engineered on command. The contribution is a
measurement and governance discipline: define the path problem clearly
enough that claims about asymmetry can be supported, rejected, or
revised.

The protocol's boundary conditions are therefore part of the evidence.
Every report should disclose benchmark choice, return frequency, sample
period, fee treatment, data source, liquidity assumptions, drawdown
threshold, and whether episodes are benchmark-defined,
portfolio-defined, or both. Capture ratios depend on the distribution of
up and down observations. Recovery burden reduction depends on episode
definition. Time underwater can improve because losses were smaller,
because recovery participation was higher, or because the benchmark was
chosen differently. These facts do not weaken the framework; they make
it falsifiable.

The most useful application will often produce mixed evidence. A
defensive strategy may reduce a slow bear-market drawdown but lag a fast
recovery. A diversifier may help in an inflation shock and fail in a
liquidity shock. A dynamic rule may improve one episode and whipsaw in
the next. Those outcomes are not excuses. They are the information the
allocator needs.

The deeper boundary is cost. A portfolio can pay for protection through
premium bleed, lower beta, turnover, tracking error, whipsaw, liquidity
constraints, complexity, or model risk. The goal is not to make
uncertainty disappear. The goal is to make the tradeoff visible enough
that institutions can choose it deliberately and stay with it when the
path becomes difficult.

\section{Conclusion}\label{conclusion}

The core claim of this paper is direct: stability is the preservation of
compounding power through adverse regimes. It is not the absence of
movement, the suppression of all volatility, or a promise that drawdowns
will be avoided. Stability is the ability to remain institutionally
functional when the path becomes difficult.

Volatility matters because it affects compounding, but volatility alone
is an incomplete abstraction. The investor-lived problem is drawdown,
recovery, and submergence. A portfolio that reduces volatility
symmetrically may look safer while failing to improve the path that
institutions actually experience. A portfolio that reduces harmful
volatility more than productive volatility may improve that path without
abandoning return participation.

Losses impose nonlinear recovery burdens. The geometry of compounding
penalizes variance. Conditional skewness and coskewness determine
whether assets help or hurt when the total portfolio is under stress.
Capture ratios translate these ideas into allocator language, while
submergence measures the duration of institutional pain. Skew
engineering brings them together into a practical design discipline.

The ambition is neither timidity nor perfection. It is staying power.
The objective is not lower volatility alone. It is a return path whose
losses, gains, and recoveries are deliberately shaped around the
institution's ability to stay invested.

\clearpage

\section*{References}\label{references}
\addcontentsline{toc}{section}{References}

\begingroup
\setlength{\leftskip}{1.5em}
\setlength{\parindent}{-1.5em}

\noindent\hspace*{-1.5em}Bailey, D. H., Borwein, J. M., López de Prado,
M., \& Zhu, Q. J. (2017). \emph{The Probability of Backtest
Overfitting}. The Journal of Computational Finance, 20(4), 39-69.
\url{https://doi.org/10.21314/JCF.2016.322}

Baker, M., Bradley, B., \& Wurgler, J. (2011). \emph{Benchmarks as
Limits to Arbitrage: Understanding the Low-Volatility Anomaly}.
Financial Analysts Journal, 67(1), 40-54.
\url{https://doi.org/10.2469/faj.v67.n1.4}

Dekhayser, J., LeMar, R., \& Engle, T. (2025). \emph{Low Volatility
Equities: Portfolio Design Matters}. Northern Trust Asset Management.
\url{https://ntam.northerntrust.com/content/dam/northerntrust/investment-management/global/en/documents/research/quantitative/low-volatility-equities-portfolio-design-matters.pdf}

Driscoll, Z., \& Giordano, L. (2024). \emph{Long Volatility Investment
Strategies Primer}. Meketa Investment Group.
\url{https://meketa.com/wp-content/uploads/2024/12/MEKETA_Long-Volatility-Investment-Strategies-Primer.pdf}

Federal Reserve Board, Federal Deposit Insurance Corporation, and Office
of the Comptroller of the Currency. (2026). \emph{Revised Guidance on
Model Risk Management}. SR 26-2 / OCC Bulletin 2026-13 / FDIC
FIL-15-2026.
\url{https://www.federalreserve.gov/supervisionreg/srletters/SR2602.htm};
\url{https://occ.gov/news-issuances/bulletins/2026/bulletin-2026-13.html};
\url{https://www.fdic.gov/news/financial-institution-letters/2026/agencies-revise-interagency-model-risk-management-guidance}

Gambhirwala, D., \& Brundage, S. (2021). \emph{Market Protection for
Turbulent Times}. PGIM Quantitative Solutions.
\url{https://www.pgim.com/content/dam/pgim/us/en/pgim-quantitative-solutions/active/documents/tbd/PGIM-Quantitative-Solutions-Market-Protection-for-Turbulent-Times.pdf}

Harvey, C. R., \& Siddique, A. (1999). \emph{Autoregressive Conditional
Skewness}. Journal of Financial and Quantitative Analysis, 34(4),
465-487. \url{https://doi.org/10.2307/2676230}

Harvey, C. R., \& Siddique, A. (2000). \emph{Conditional Skewness in
Asset Pricing Tests}. The Journal of Finance, 55(3), 1263-1295.
\url{https://doi.org/10.1111/0022-1082.00247}

Hurst, B., Ooi, Y. H., \& Pedersen, L. H. (2017). \emph{A Century of
Evidence on Trend-Following Investing}. The Journal of Portfolio
Management, 44(1), 15-29.
\url{https://doi.org/10.3905/jpm.2017.44.1.015}

Kaminski, K. M., \& Zhao, Y. (2023). \emph{Crisis or Correction:
Managing Expectations for Managed Futures and Crisis Alpha}.
AlphaSimplex.
\url{https://www.alphasimplex.com/assets/files/2023.08---crisis-or-correction---kaminski-and-zhao.pdf}

Kelly, J. L., Jr.~(1956). \emph{A New Interpretation of Information
Rate}. Bell System Technical Journal, 35(4), 917-926.
\url{https://doi.org/10.1002/j.1538-7305.1956.tb03809.x}

López de Prado, M. (2018). \emph{Advances in Financial Machine
Learning}. Wiley.
\url{https://www.wiley.com/en-us/Advances+in+Financial+Machine+Learning-p-9781119482086}

Markowitz, H. (1952). \emph{Portfolio Selection}. The Journal of
Finance, 7(1), 77-91.
\url{https://doi.org/10.1111/j.1540-6261.1952.tb01525.x}

Mauboussin, M., \& Callahan, D. (2025). \emph{Drawdowns and Recoveries:
Base Rates for Bottoms and Bounces}. Morgan Stanley Investment
Management.
\url{https://www.morganstanley.com/im/publication/insights/articles/article_drawdownsandrecoveries.pdf}

Moreira, A., \& Muir, T. (2017). \emph{Volatility-Managed Portfolios}.
The Journal of Finance, 72(4), 1611-1644.
\url{https://doi.org/10.1111/jofi.12513}

Pak, E. (2014). \emph{What Are Upside and Downside Capture Ratios?}
Morningstar.
\url{https://global.morningstar.com/en-ca/personal-finance/what-are-upside-and-downside-capture-ratios}

Rockafellar, R. T., \& Uryasev, S. (2000). \emph{Optimization of
Conditional Value-at-Risk}. Journal of Risk, 2(3), 21-41.
\url{https://doi.org/10.21314/JOR.2000.038}

Rook, D., Golosovker, D., \& Monk, A. (2023). \emph{Submergence =
Drawdown Plus Recovery}. SSRN.
\url{https://doi.org/10.2139/ssrn.4346463}

Van Hemert, O., Ganz, M., Harvey, C. R., Rattray, S., Sanchez Martin,
E., \& Yawitch, D. (2020). \emph{Drawdowns}. The Journal of Portfolio
Management, 46(8), 34-50. \url{https://doi.org/10.3905/jpm.2020.1.170}

Xiang, V., Ben-Akiva, O., \& Wee, J. (2023). \emph{Not All Low
Volatility Portfolios Are The Same}. Man Group.
\url{https://www.man.com/insights/not-all-volatility-portfolios}

\endgroup

\end{document}